\documentclass[twocolumn,           
               showpacs,            
               preprintnumbers,     
               aps,                 
               prd,                 %
               superscriptaddress,   
               nofootinbib,         
               tightenlines,        
               floats,floatfix      
               ]{revtex4}

\usepackage{amsmath}
\usepackage{color}
\usepackage{amssymb}
\usepackage{graphicx}

\begin{document}


\title{Accretion of phantom scalar field into a black hole}


\author{J. A. Gonz\'alez}
\affiliation{Instituto de F\'{\i}sica y Matem\'{a}ticas, Universidad
              Michoacana de San Nicol\'as de Hidalgo. Edificio C-3, Cd.
              Universitaria, 58040 Morelia, Michoac\'{a}n,
              M\'{e}xico.}

\author{F. S. Guzm\'an}
\affiliation{Instituto de F\'{\i}sica y Matem\'{a}ticas, Universidad
              Michoacana de San Nicol\'as de Hidalgo. Edificio C-3, Cd.
              Universitaria, 58040 Morelia, Michoac\'{a}n,
              M\'{e}xico.}


\date{\today}


\begin{abstract}
Using numerical methods we present the first full nonlinear study of 
phantom scalar field accreted into a black hole. 
We study different initial 
configurations and find that the accretion of the field into the black 
hole can reduce 
its area down to 50 percent within 
time scales of the order of few masses of the initial horizon.
The analysis includes the cases where the total energy of the 
space-time is positive or negative. 
The confirmation of this effect in full nonlinear general relativity 
implies that the accretion of exotic matter could be considered an evaporation process.
We speculate that if this sort of exotic matter has some 
cosmological significance, this black hole area reduction 
process might have played a crucial role in black hole formation 
and population.
\end{abstract}




\maketitle


The study of black holes in general relativity as thermodynamical
systems is an extensive area of research.
One of the most interesting features of black holes that can 
be studied, is the behavior of the area of the horizon during 
the process of accretion of matter.
Assuming that the null energy condition is satisfied, the 
area of the horizon can only increase \cite{HawkingEllis-Book}. 
If this condition
is violated then ``strange'' things can happen, for example, 
the area of the horizon could decrease and the black hole could shrink.
This scenario could be astrophysically and cosmologically interesting
if some evidences of exotic matter, violating the null energy 
condition would be found.  For instance, nowadays standard cosmological models,
based on observations allow that matter with these types of 
properties might well play the role of dark energy \cite{darkenergy}.

In this article we present the accretion of a phantom scalar field into a black
hole in the full nonlinear case using numerical methods.
The model of phantom scalar field is based on the following Lagrangian density 

\begin{equation}
{\cal L} = {\cal R} + \frac{1}{2} g_{\mu\nu} \partial^{\mu}\phi \partial^{\nu} 
\phi - V(\phi),\label{eq:Lagrangian}
\end{equation}

\noindent where ${\cal R}$ is the Ricci scalar of the space-time, 
$g_{\mu\nu}$ is the space-time metric, $\phi$ is the scalar field and $V(\phi)$ its
potential, which we choose to be $V(\phi)=\lambda \phi^2$
with $\lambda$ a given constant. The difference between the phantom and the
standard scalar field is the sign in the kinetic term in the Lagrangian. 

When the action constructed 
with such a Lagrangian density is varied with respect to the metric, the 
arising Einstein's equations are related to the stress energy-tensor
$T_{\mu\nu} = -\partial_{\mu} \phi \partial_{\nu}\phi
-\frac{1}{2}g_{\mu\nu}[-\partial^{\alpha}\phi \partial_{\alpha}\phi + 2V]$.
Such a stress-energy tensor
violates the null energy condition $T_{\mu\nu}k^{\mu}k^{\nu} \le 0$, 
where $k^{\mu}$ is a null vector. The immediate implication of this 
violation is the violation also of the weak energy condition, 
which in turn implies that observers following timelike trajectories 
might measure negative energy densities. 

The fact that a phantom scalar field violates the null energy condition 
motivates the study of possible unusual implications in astrophysical 
scenarios, because the area increasing theorem does not apply in this 
case \cite{HawkingEllis-Book}. 

Previous experience shows that such a scalar field with $V=0$ is able to 
produce an area reduction effect, as found during the evolution of wormholes 
supported by a scalar field that collapse to form black holes 
\cite{whs}. Inspired by such an area reduction process we 
explore in this article the full nonlinear accretion of a phantom scalar 
field, that is, the case where a nonzero potential is involved.

{\it The evolution equations}. 
We formulate Einstein's equations coupled to the phantom scalar 
field in a suitable form for the numerical integration of spherically 
symmetric black hole solutions. We use geometrized units for which the 
speed of light and Newton's constant are equal to one. 
We write the metric in the general form for the coordinate system 
$(t,r,\theta,\varphi)$ in spherical coordinates:

\begin{eqnarray}
ds^2 &=& -\left( \alpha^2 - \beta^r \beta^r \frac{g_{rr}}{\chi} \right)dt^2 
	 + 2\beta^r \frac{g_{rr}}{\chi}dtdr  \nonumber\\
	&+& \frac{1}{\chi} \left[g_{rr} dr^2 + g_{\theta\theta}
	(d\theta^2 + \sin^2 \theta d\varphi^2)\right], \label{eq:metric}
\end{eqnarray}

\noindent where $\chi$ acts as a conformal factor relating this 
metric to a spacelike flat metric, $\beta^r$ is the only nonzero 
component of the shift vector and $\alpha$ is the lapse function.

The procedure used to solve Einstein's equations is based on 
a 3+1 decomposition of the space-time and the solution of an initial 
value problem assuming a puncture type of initial data \cite{Bruegmann}. 
In order to follow the evolution of the initial data we 
found appropriate the use of the Generalized BSSN (GBSSN) evolution 
system of 
equations defined previously for spherical symmetry in \cite{Brown} as 
opposed to previous successful analyses related to the accretion of 
a scalar field using Eddington-Finkelstein coordinates under the 
Arnowitt-Deser-Misner (ADM) formulation that 
requires excision \cite{Thornburg}. 

The evolution equations in the GBSSN system for the spherically 
symmetric space-time described by (\ref{eq:metric}) reduce to a set 
of evolution equations for the conformal factor $\chi$, the conformal metric 
functions $g_{rr}$ and $g_{\theta\theta}$, the nonzero trace-free part 
of the conformal extrinsic curvature $A_{rr}$, the trace of the 
extrinsic curvature $K$ and the contracted conformal Christoffel nonzero symbol 
$\Gamma^{r}$. The explicit expression for the evolution of these 
variables can be found in \cite{Brown,BShapiro}.

The remaining equations of the geometry involve the evolution of the 
gauge. For the lapse we choose the 1+$\log$ slicing condition 
$\partial_t \alpha = \beta^a \partial_a 
\alpha - 2\alpha K$, and for the shift we implemented the recipe 
for the $\Gamma$-driver condition 
$\partial_t \beta^a = \frac{3}{4}B^a + \beta^c \partial_c \beta^a$ and
$\partial_t B^a = \partial_t \Gamma^a + \beta^c \partial_c B^a
        - \beta^c \partial_c \Gamma^a - \eta B^a$, 
which helps to avoid the slice 
stretching near the horizon which is known to kill the numerical 
evolution. In order to avoid instabilities near the puncture we 
implement a sort of excision without excision using a factor function 
on the sources of the evolution equations of the form $(r/(1+r))^4$ 
from the coordinate origin out to one quarter of the size of the 
apparent horizon radius. Even though this function violates the
constraints, the convergence tests show that such violation does
not propagate outside of the black hole horizon. 

The evolution of the scalar field is given by the Klein-Gordon 
equation

\begin{equation}
\Box \phi = \frac{1}{\sqrt{-g}}\partial_{\mu}
	[\sqrt{-g}g^{\mu\nu} \partial_{\nu}\phi] = -\partial_{\phi}V,
\label{eq:kg}
\end{equation}

\noindent where $g$ is the determinant of the space-time metric. We solve 
this equation as a set of two equations for two first-order variables,
$\pi=\partial_t \phi$ and $\xi = \partial_r\phi$, coupled to the 
evolution of the geometry of the space-time.

{\it Initial data.} 
In order to add the contribution of the scalar field to Einstein's 
equations, it is necessary to solve the constraints at the initial time 
slice which we assume for simplicity to be time-symmetric, which 
in turn implies that the momentum constraint is identically satisfied. 
The scalar field contribution consists of a scalar field pulse which is used 
to solve the Hamiltonian constraint. Another 
important ingredient is that we set 
a space-time ansatz similar to that of a Schwarzschild black hole 
in isotropic coordinates. Thus we assume that the metric at initial time
has the form

\begin{equation}
g_{rr} = 1, ~~~
g_{\theta \theta} = r^2, ~~~
\chi = \left(1 + \frac{M}{2r} + u\right)^{-4},
\label{eq:metric_id}
\end{equation}

\noindent where $M$ is the ADM mass at the puncture and agrees within
numerical precision with the mass of the apparent horizon and
$u=u(r,t=0)$ is the function to be determined through the solution of 
the Hamiltonian constraint. The Hamiltonian constraint in terms of $u$ reads

\begin{equation}
\partial_{rr} u = 
	\frac{1}{8} (\partial_r \phi)^2 \chi^{-1/4}
  	-\frac{2 \partial_r u}{r} - \frac{V}{4}\chi^{-5/4},
\label{eq:u}
\end{equation}

\noindent which we solve using an ordinary differential equation integrator.
We finally rescale $r$ in such a way that $\chi \rightarrow 1$ 
when $r \rightarrow \infty$.
We notice that with this procedure it is also possible to construct initial data 
consisting of a large enough amount of scalar field so that the 
ADM mass could become negative.

It only remains to choose the gauge at initial time, for which we use an 
initially precollapsed lapse $\alpha=(1+M/2r)^{-2}$, zero shift 
$\beta^r=0$ and zero time derivative $B^r=0$. 
Because the initial time slice has been 
chosen to be time-symmetric, the extrinsic curvature components are all 
set to zero initially.
These data are used to start up a Cauchy evolution using the 
GBSSN evolution equations, the gauge conditions 
and the Klein-Gordon equation.

{\it Numerical methods.}
The numerical method used to approximate the 
constraint and evolution equations is a second-order finite differences 
approximation. We only perform the evolution on a finite domain with 
artificial boundaries at a finite value of the radial coordinate. 
The integration in time uses a method of lines with a fourth-order 
accurate Runge-Kutta time integrator. Throughout the evolution we 
monitor the Hamiltonian, Momentum and Gamma constraints \cite{Brown}. 
We check that they converge with second order to zero as the resolution 
is increased.

{\it Apparent horizon location}. As we are interested in tracking the
behavior of the black hole horizon, a useful diagnostics tool is the
location of the apparent horizon during the evolution.
We locate it from the definition
of a marginally trapped surface (MTS) through the condition
$\Theta = \nabla_i n^i + K_{ij} n^i n^j - K =0$, where $n^i$ is an 
outward pointing unit vector normal to the apparent horizon, $K_{ij}$ are the 
components of the extrinsic curvature of the spacelike hypersurface 
on which one calculates the MTSs, and $K$ its trace.
The apparent horizon 
is the outermost MTS. For the metric (\ref{eq:metric}) this equation is 
written as

\begin{eqnarray}
\Theta &=& \frac{1}{g_{\theta\theta}\sqrt{\chi g_{rr}}}
	\left(\chi (\partial_r g_{\theta\theta}
	)  - g_{\theta\theta}  (\partial_r \chi)\right) \nonumber\\
	&+& 2 \left( \frac{A_{rr}}{2g_{rr}} - \frac{1}{3} K \right).
\label{eq:ah}
\end{eqnarray}

\noindent In order to track the evolution of the apparent horizon we 
calculate $\Theta$ at every time step and locate the outermost zero of 
it at the coordinate radius $r_{AH}$ and calculate the mass of this 
horizon $M_{AH} = R_{AH}/2$, where $R_{AH}=\sqrt{g_{\theta\theta}/\chi}$ 
is the areal radius evaluated at $r_{AH}$.

{\it Mass.} In order to have an estimate of the space-time mass we
use the Misner-Sharp mass defined by \cite{MisnerSharp}

\begin{eqnarray}
M_{MS} &=& \frac{R}{2}
\left[ 1 + \frac{1}{\alpha^2}(\partial_t R)^{2}  \right.
-2\frac{\beta^r}{\alpha^2}(\partial_t R)(\partial_r R) \nonumber\\
&-& \left. \left( \frac{\chi}{g_{rr}} -\frac{(\beta^r)^2}{\alpha^2} \right)
(\partial_r R)^{2} \right], \label{eq:MisnerSharpMass}
\end{eqnarray}

\noindent where 
$R=\sqrt{g_{\theta\theta} / \chi}$ is the areal radius.

We estimate the ADM mass as 
$M_{ADM} = \lim_{r \to \infty} M_{MS}$ so that we can estimate the mass of 
the space-time at any time during the evolution even though we know 
it should be constant in time.

{\it Null rays and event horizon}. We track a bundle of radial outgoing null rays 
in order to i) approximately locate the event horizon as the surface 
that separates outgoing null rays that reach spatial infinity from those that do not when launched toward the future and ii) certify that the light-cones point in the correct 
direction at the apparent and event horizons.

\begin{figure}[ht]
\includegraphics[width=8cm]{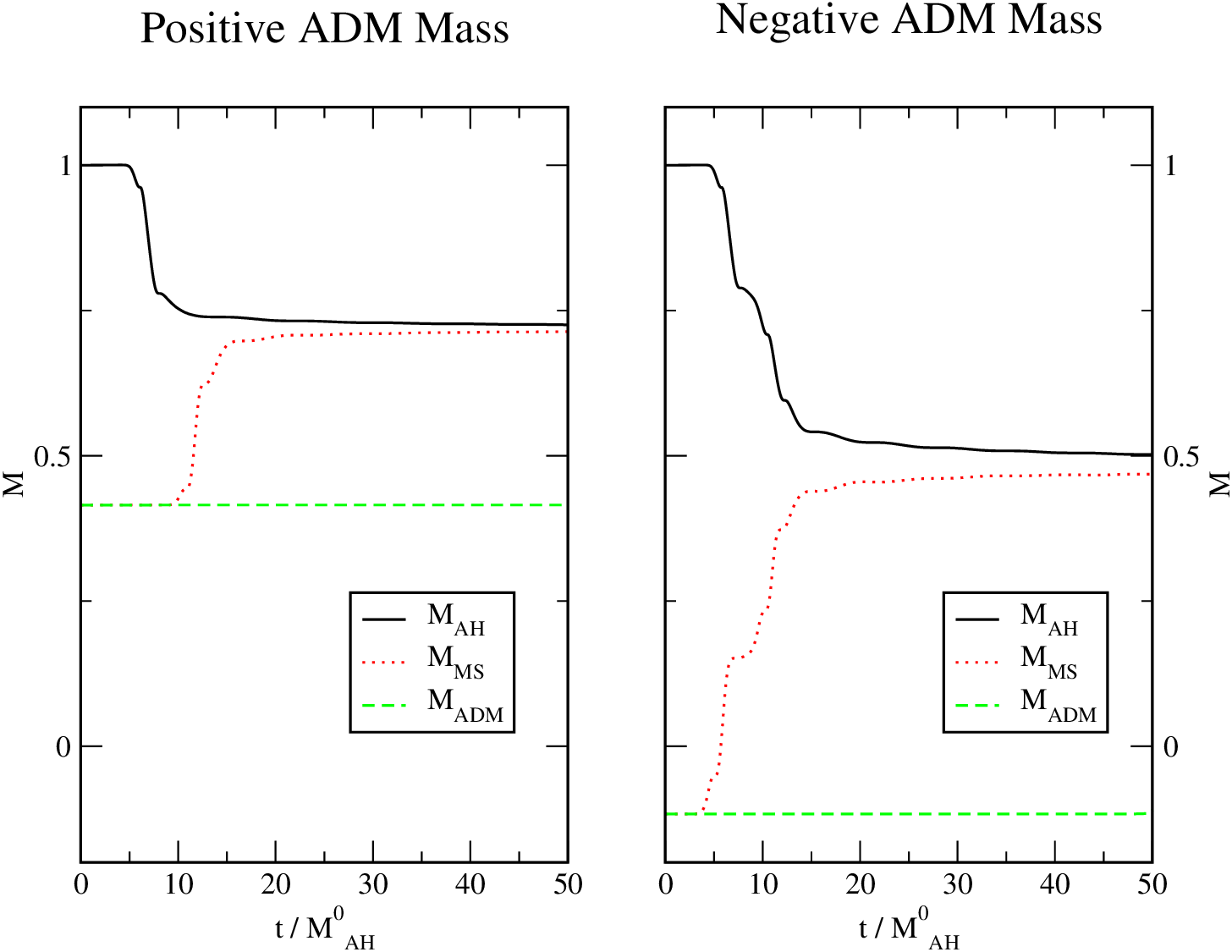}
\caption{We show the evolution of different masses 
versus the coordinate time. The solid line represents the mass of
the apparent horizon, the dotted line the Misner-Sharp mass extracted at
$r\sim 15M^{0}_{AH}$ and the dashed line shows the ADM mass of the space-time. 
In the left panel the ADM mass of the space-time is positive and 
in the right one it is negative. In both scenarios the apparent horizon mass 
decreases and approaches to the Misner-Sharp mass. In the first case
the black hole loses approximately $25\%$ of its mass while in the second
case it losses almost half of the mass.}
\label{fig:AHradius_case}
\end{figure}

{\it Results}.
We analyze different cases with various parameters of the initial scalar field 
profile and different values of $\lambda$ including the case $\lambda=0$ and obtain
qualitatively similar 
results. We use a Gaussian profile or a train of successive Gaussians with different amplitudes and widths 
for the scalar field. In particular we present two cases characterized by 
positive and negative ADM masses \cite{comentario}. In both cases $\lambda=-0.1$; in the 
case of Lagrangian 
(\ref{eq:Lagrangian}) a negative potential with a maximum is appropriate to keep the 
scalar field stable. The 
results are shown in
Fig. \ref{fig:AHradius_case}, where we show the evolution 
of the apparent horizon, Misner-Sharp and ADM masses \cite{MisnerSharp}.
In both cases the area of the apparent horizon decreases and asymptotically in time 
converges to the Misner-Sharp mass. It is worth noticing that for the case of negative ADM mass
we used an initial train of scalar field Gaussians that are being successively accreted, 
which produces the step-like mass decrease of the horizon mass observed in Fig. 1.

\begin{figure}[ht]
\includegraphics[width=8cm]{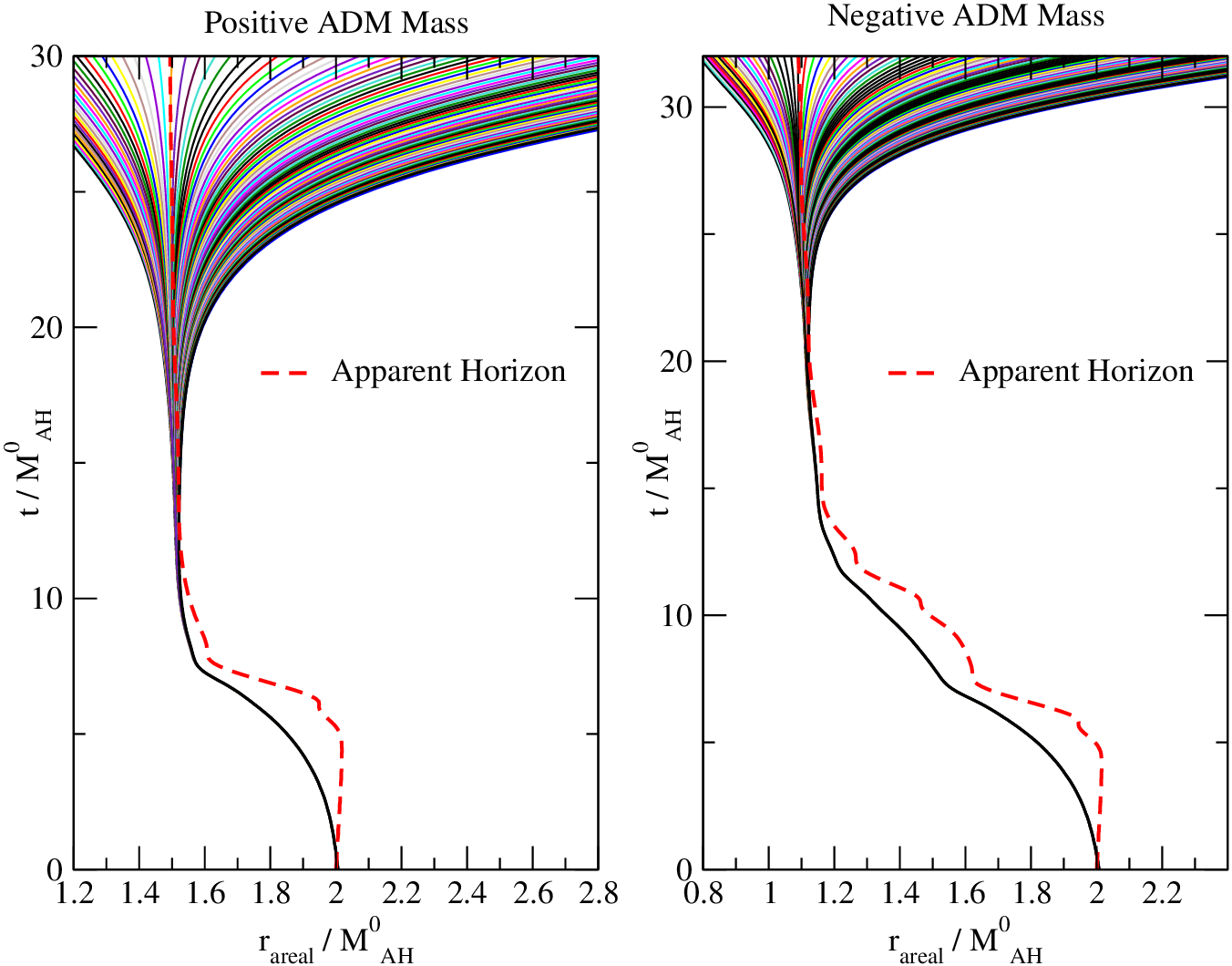}
\caption{We present a bundle of outgoing null rays showing the 
approximate location of the event horizon for the runs shown in 
Fig. \ref{fig:AHradius_case}. It can be seen that 
this horizon decreases so as the apparent horizon does, showing that our 
results are not slice dependent. It is interesting to notice that 
the apparent horizon lies outside the event horizon.}
\label{fig:nullrays}
\end{figure}

In order to show that the decrease of the black hole area is not a gauge artifact, 
in Fig. \ref{fig:nullrays} we show a fine-tuned bundle of outgoing 
null rays for the two cases, and show the approximate 
location of the event horizon. The fact that the event horizon decreases with the 
accretion of the scalar field is a strong evidence that the hole area 
is truly decreasing since the event horizon is a gauge invariant 
surface unlike the solely consideration of the apparent horizon.

\begin{figure}[ht]
\includegraphics[width=8cm]{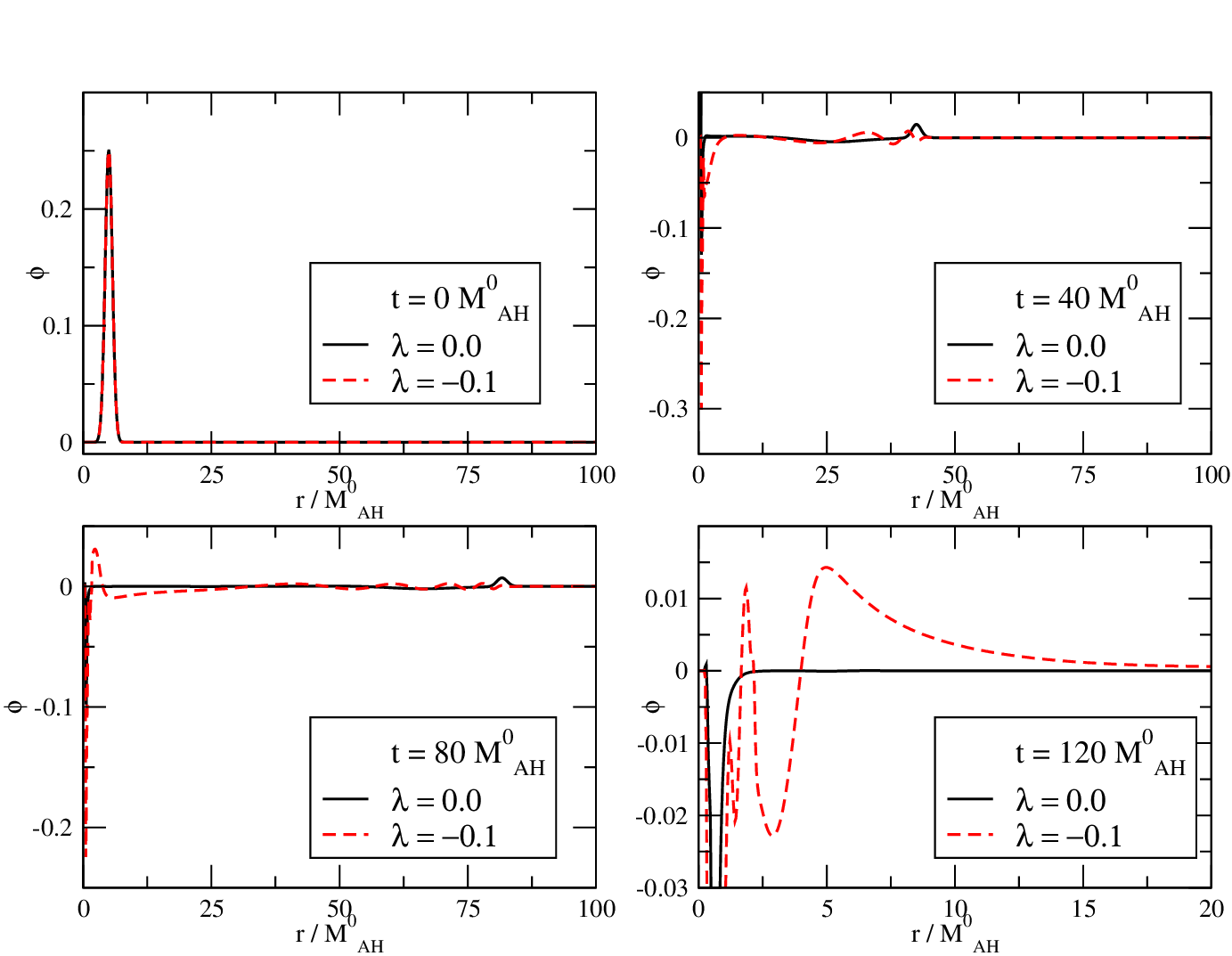}
\caption{We show snapshots of the scalar field at various 
times for $\lambda=0$ (continuous line) and $\lambda=-0.1$ (dashed line).
The initial profile of the scalar field is the same, although the energy density is 
slightly different for both cases, nevertheless the time scales are 
comparable. We point out various effects: i) there is an initial outgoing pulse of scalar 
field due to the time-symmetry of the initial data in both cases, ii) however in the 
$\lambda=0$ case 
the amplitude of the field quickly stabilizes around zero in the domain left behind by the 
outgoing pulse, whereas the field does not stabilize in the nonzero case, and instead new 
pulses of the scalar field are being ejected from the region near the horizon.}
\label{fig:ghost_vs_phantom}
\end{figure}

Another interesting result is that the accretion rate and 
localization of the scalar field depends on the scalar field potential. 
As an example, in Fig. \ref{fig:ghost_vs_phantom} we compare the behavior
of the scalar field in the case where the potential is nonzero with the case
of zero potential.
In the case of the zero potential the scalar field is quickly 
accreted whereas in the other case the scalar field gets packed near the 
horizon and some scalar field is radiated away.

\begin{figure}[ht]
\includegraphics[width=8cm]{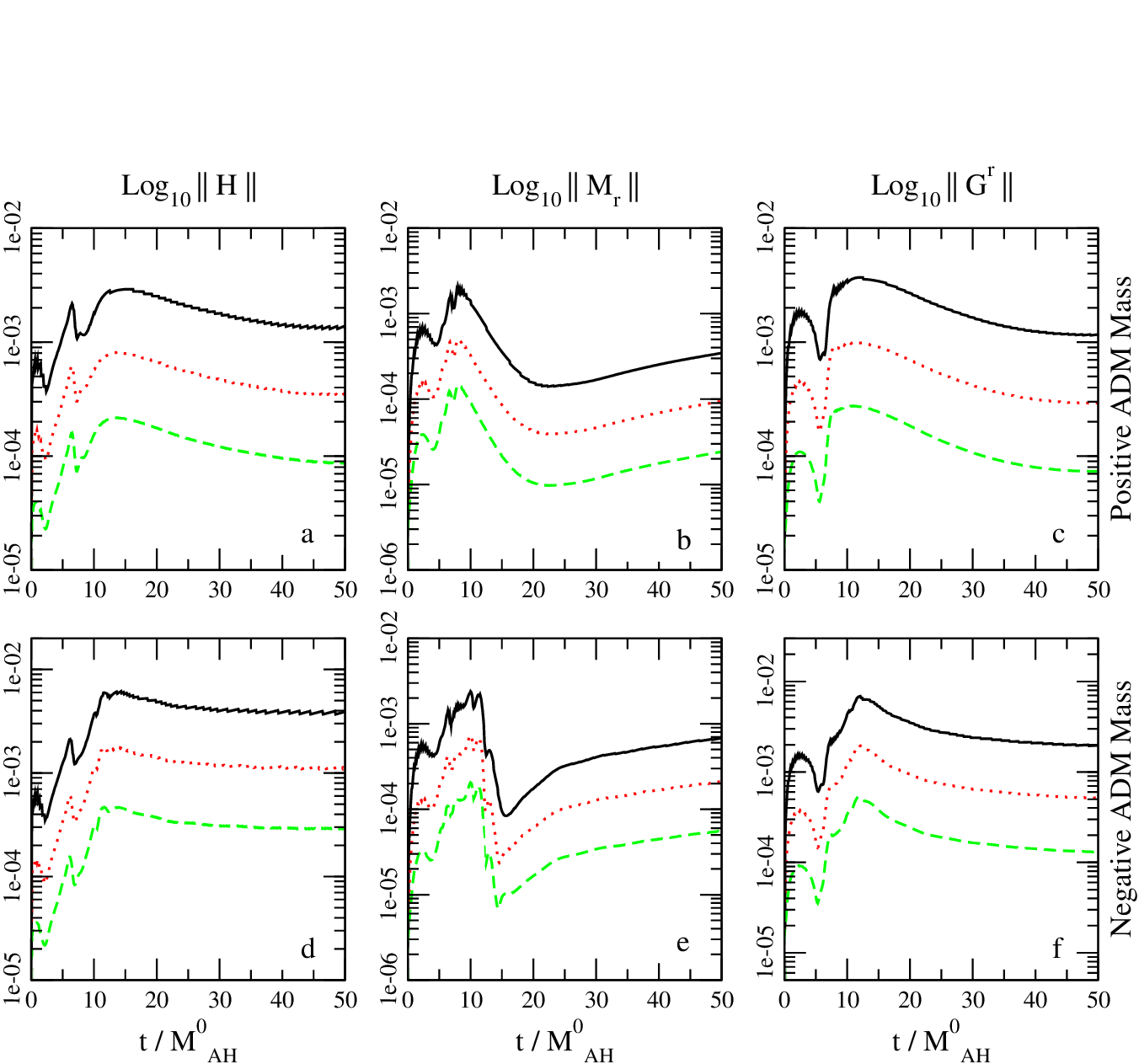}
\caption{We show the convergence of the $L_2$ norm of the 
violation of the Hamiltonian, momentum and $\Gamma^r$ constraints to zero.
The plots a,b and c correspond to the simulation with positive ADM mass and
the plots d,e and f to the negative ADM mass case.}
\label{fig:convergence}
\end{figure}

Finally, in order to validate our results, we have verified the convergence 
of our results to second order so as the convergence of the violation 
of the constraints to zero. In Fig. \ref{fig:convergence} we show the 
convergence of the $L_2$ norm of the 
constraint violations to zero for the simulations presented in Fig. \ref{fig:AHradius_case}.

{\it Discussion.}
In order to illustrate the process we consider the hypothetical case 
of a black hole  space-time with 
positive ADM mass. 
Assuming the mass of the original black hole apparent horizon is 
$M_{phys} = 10^9 M_{\odot}$, and that a pulse of scalar field with amplitude $0.25$ 
is accreted, the time-scale for the field to reduce the mass of the hole in 25\%  is 
$t \sim 1$day in coordinate time (see the results in Fig. \ref{fig:AHradius_case}).

In theory, due to the fact that our results scale with the initial
mass of the black hole, it would always be possible to construct the
appropriate scalar field profile so that the process of accretion 
could be iterated reducing half of the mass of the horizon each 
time.

The results presented constitute solid steps toward exploring the accretion of
phantom scalar field dark energy under different astrophysical
conditions, mainly those related to cosmologically motivated
potentials and wavelengths of the scalar field \cite{darkenergy}. It would also 
be interesting to investigate the full nonlinear accretion of phantom energy in
black holes immersed in a FRW geometry like those in Mc Vittie-type
black hole solutions \cite{Faraoni2007}, because these scenarios could
have cosmologically interesting implications. 

One of the properties of this type of field in the cosmological scenario is that 
if $T_{\mu\nu}$ is the stress-energy tensor, the 
ratio between the radial pressure $p_r = n^{\mu}T_{\mu r}$ and the 
energy density $\rho = n^{\mu}n^{\nu}T_{\mu \nu}$ measured by Eulerian observers with $n^{\mu}$ 
the normal to the spatial hypersurfaces, provides an equation 
of state such that $\omega_r = \frac{p_r}{\rho} < -1$ 
in the case of nonzero potential, and $\omega_r = 1$ in the case of zero 
potential.
These properties change when the configurations 
depend on the spatial coordinates and the potential is nonzero. For instance, in the 
case discussed here, the scalar field depends on the spatial coordinate and 
there are regions where the equation of state 
$\omega < -1$ is satisfied, 
however there are also regions where $\phi$ approaches zero and the potential becomes zero too, in 
which case the scalar field would behave as stiff matter $\omega=1$.  

These results should imply important restrictions on the parameters 
of phantom dark energy models related to the abundance and mass of 
black holes and the abundance and lifetime of PBHs for instance. 
We also expect that the type of process shown in this article competes 
with other horizon-reducing processes of black holes.

A detailed study containing various initial conditions, more violent 
horizon-shrinking processes, dynamical properties of horizons and the 
study of the possibility of producing a naked the singularity is in progress 
\cite{detailedpaper}.


\acknowledgments

We thank A. Corichi, C. L\'opez, O. Sarbach and T. Zannias 
for many stimulating discussions. 
This work was supported in part by grants 
CIC-UMSNH 4.9 and 4.23,
PROMEP UMICH-PTC-210, UMICH-CA-22 from SEP Mexico,
COECyT Michoacan S08-02-28 and 
CONACyT grant numbers 79601 and 79995.



\begin{thebibliography}{}
\bibitem{comentario}
	It is not surprising that it is possible to construct space-times
	with negative ADM mass because the contribution of the scalar field to 
	the energy density is negative.
\end{thebibliography}
\end{document}